\documentclass[epj,final]{svjour}
\usepackage{graphicx,amssymb}
\usepackage[english]{babel}

\newcommand{\debhu}{Debye-H\"{u}ckel}
\newcommand{\kb}{\mathrm{k_B}}

\begin{document}

\title{Non-extensive resonant reaction rates in astrophysical plasmas}

\author{F.~Ferro\inst{1}\and A.~Lavagno\inst{1,2} 
\and P.~Quarati\inst{1,3}
\thanks{\email{piero.quarati@polito.it}}
}

\institute{Dipartimento di Fisica, Politecnico di Torino, I-10129 Torino, Italy
\and INFN - Sezione di Torino, I-10125 Torino, Italy
\and INFN - Sezione di Cagliari, I-09042 Monserrato, Italy}
\date{Received: date / Revised version: date}
\abstract{We study two different physical scenarios of thermonuclear reactions in stellar plasmas proceeding through a
narrow resonance at low energy or through the low energy wing of a wide resonance at high energy. Correspondingly, we
derive two approximate analytical formulae in order to calculate thermonuclear resonant reaction rates inside very
coupled and non ideal astrophysical plasmas in which non-extensive effects are likely to arise. Our results are
presented as simple first order corrective factors that generalize the well known classical rates obtained in the
framework of Maxwell-Boltzmann statistical mechanics. As a possible application of our results, we calculate the
dependence of the total corrective factor with respect to the energy at which the resonance is located, in an extremely
dense and non ideal carbon plasma.
} 

\maketitle

\section{Introduction}\label{intro}
Cussons, Langanke and Liolios~\cite{Langanke} proposed, on the basis of experimental measurements at energy $E\sim
2.4\,\mathrm{MeV}$, that the resonant behavior of the stellar $^{12}\mathrm{C}+{^{12}\mathrm{C}}$ fusion cross
section could continue down to the astrophysical energy range, thus leading rise, beside the usual \debhu\/
screening~\cite{Salpeter} (whose corrective factor is $f_{\mathrm{S}}$), to further plasma resonant screening effect
(conventionally described by a suitable $f_{\mathrm{RS}}$ factor). While $f_{\mathrm{S}}>1$ enhances the reaction rate,
it has been pointed out that $f_{\mathrm{RS}}<1$, {\it i.e.} the resonant screening effect is likely to reduce the
rate. The reduction of the resonant rate due to resonant screening correction amounts to 11 orders of magnitude at the
resonant energy of $400\,\mathrm{keV}$, influencing the carbon ignition density in white dwarfs. Itoh {\it et
al.}~\cite{Japan} have shown that, using an effective screening potential obtained by one-component-plasma (OCP) Monte
Carlo experiments, the overall effect does indeed strongly enhance the carbon-carbon reaction rate by a considerable
amount ({\it i.e.} $f_{\mathrm{T}}\equiv f_{\mathrm{S}}\cdot f_{\mathrm{RS}}\gg 1$) 
because of the global screening domination
that amounts to an enhancement of the rate by 12 orders of magnitude, with important implications for hydrostatic
burning in carbon white dwarfs. Given the $^{12}\mathrm{C}+{^{12}\mathrm{C}}$ reaction, the current
hypothesis~\cite{Langanke} is that the entrance channel width is much smaller than the total resonance width, the
latter being much smaller than the resonance energy. The same picture could possibly apply to other fusion reactions
between medium-weighted nuclei ({\it e.g.} to the $^{16}\mathrm{O}+{^{16}\mathrm{O}}$ reaction, which could be an
active burning stage in some white dwarfs~\cite{white dwarf}).

The previous discussion refers to extremely dense stellar plasmas, characterized by a temperature $T\sim
10^8\,\mathrm{K}$, a mass density $\rho\sim 10^9\,\mathrm{g\cdot cm^{-3}}$ and a plasma parameter $\Gamma<178$. In
these physical conditions, we expect that non-extensive effects could also arise. We briefly recall that such plasmas
show deviations from the several assumptions that are the basis of Maxwell-Boltzmann distribution. Long range many-body
nuclear correlations and memory effects, among others, can be sufficient to justify the use of a distribution function
which slightly deviates from the standard Maxwell-Boltzmann one~\cite{LavQua} (see also, for example, ref.~\cite{KLLQ}
for a discussion on physical conditions in which non-extensivity needs to be taken into account). Our aim is to derive
a simple first order formula in order to express non-extensive corrections for reactions proceeding through narrow
resonances: we will consider this case in subsection~\ref{narrow NE r}.

In addition to the above resonant reactions in white dwarfs, let us mention few resonant reactions occurring in the
stellar interior (like the Sun interior), where no carbon burning is active. It is well known that, along with the
reactions of the proton-proton chain, many reactions of the CNO cycle are presently believed to be non-resonant at the
relevant stellar energies around the maximum $E_{0}$ of the Gamow peak (whose order of magnitude is $E_0\sim
10\,\mathrm{keV}$ inside the Sun's core). However, the $^{17}\mathrm{O}(p,\alpha )^{14}\mathrm{N}$ reaction, which
belongs to the CNO-II subcycle, shows a very narrow resonance ($^{18}\mathrm{F}$ being the compound nucleus) located at
an energy between $65\,\mathrm{keV}$ and $75\,\mathrm{keV}$, with a proton partial width $\Gamma _{p}\simeq
22\,\mathrm{neV}$; this might be taken into account when solar model calculations are carried out
(see~\cite{BahcallExperimental} and references therein). Moreover, one could actually point out that many non-resonant
reactions in the CNO cycle proceed indeed through the low energy wing of a wide resonance located at energy
$E_{\mathrm{R}}\gg E_0$, as was already observed by Clayton~\cite{Clayton} (among others, we recall also the
$^{14}\mathrm{N}(p,\gamma)^{15}\mathrm{O}$ reaction; its $S$ factor has been very recently measured by~the LUNA group
\cite{Formicola}). In this case, we could adopt a very reliable resonant formalism that allows us to evaluate the
astrophysical factor of the reaction at the Gamow peak energy, $S(E_0)$, by relying exclusively on the experimental
data of the resonance at the relatively high energy $E_{\mathrm{R}}$.

For the sake of completing this discussion, let us consider the $^{12}\mathrm{C}(p,\gamma)^{13}\mathrm{N}$ reaction
(CNO-I subcycle). At its Gamow energy in the Sun, $E_0=24.68\,\mathrm{keV}$, this reaction seems to be
non-resonant~\cite{Rolfs}, but it shows a wide resonance at energy $E_{\mathrm{R}}=424\,\mathrm{keV}\gg E_0$, with a
total width $\Gamma_{\mathrm{T}}\simeq 40\,\mathrm{keV}$, and an electromagnetic channel width $\Gamma_{\gamma}\simeq
0.77\,\mathrm{eV}$. It is an easy matter to express the astrophysical factor $S(E_0)$ by means of $E_{\mathrm{R}}$,
$\Gamma_{\mathrm{T}}$ and $\Gamma_{\gamma}$ only, through a Breit-Wigner approximation formula~\cite{Hussein}.

As it has been pointed out (see, for example, refs.~\cite{Clayton paper,CKLLMQ}) a very slight non-extensive
deformation could also arise in the Sun interior: we will derive a first order non-extensive corrective factor for
nuclear reactions proceeding through the low energy wing of a wide resonance in subsection~\ref{wide NE R}.

From the previous discussion it follows that it is useful to develop an analytical theory dealing with reactions that
proceed through narrow or wide resonances; this work was already achieved in ref.~\cite{Clayton}. In this paper we
extend the known classical results to the case in which non-extensive corrections arise, and we present a possible
application to the $^{12}\mathrm{C}+{^{12}\mathrm{C}}$ reaction in a white dwarf's plasma.

\section{Classical resonant reaction rates}
In this section, we briefly review some of the known results about the analytical reaction rate calculations in
astrophysical plasmas, that were obtained in the framework of the classical Maxwell-Boltzmann statistics
(MB)~\cite{Clayton,Rolfs}.

Let us consider a thermonuclear reaction between two nuclei $i$ and $j$. We define the classical reaction rate,
$r_{ij}^{\mathrm{MB}}$, by
\begin{eqnarray}
r_{ij}^{\mathrm{MB}}&=&\frac{N_i N_j}{1+\delta_{ij}}\langle v_{ij}\sigma_{ij}\rangle_{\mathrm{MB}}=\nonumber\\
&=&\frac{N_i N_j}{1+\delta_{ij}} \int_{0}^{+\infty}\phi_{\mathrm{MB}}(E)v_{ij}(E)\sigma_{ij}(E)\mathrm{d} E\; ,
\label{reaction rate def}
\end{eqnarray}
where $N_i$, $N_j$ are the particle densities, $E$ is the relative energy (in the center of mass frame of reference) at
which the reaction occurs, $v_{ij}$ is the relative velocity between two fusing nuclei, $\sigma_{ij}$ is the reaction
cross section, $\phi_{\mathrm{MB}}$ is the Maxwell-Boltzmann energy distribution function, and $\delta_{ij}=1$ if
nuclides $i$ and $j$ are identical, $\delta_{ij}=0$ otherwise.

As far as resonant reactions are concerned, we can distinguish two different cases of physical interest. Let be
$E_{\mathrm{R}}$ the resonance energy, $E_0$ the Gamow energy and $\Gamma_{\mathrm{T}}$ the total resonance width. If
the following conditions are satisfied,
\begin{equation}
\cases{E_{\mathrm{R}}\approx E_0\cr\Gamma_{\mathrm{T}}\ll E_{\mathrm{R}}\cr},\label{narrow res low energy}
\end{equation}
the reaction proceeds through a narrow resonance at low energy (this case will be labelled as ``r''). The
$^{12}\mathrm{C}+{^{12}\mathrm{C}}$ reaction occurring inside a white dwarf, beside the hypotheses already discussed in
sect.~\ref{intro}, belongs to this first scenario.

On the contrary, if the following inequalities are satisfied,
\begin{equation}
\cases{E_{\mathrm{R}}>E_0\;\mathrm{or}\; E_{\mathrm{R}}\gg E_0\cr\Gamma_{\mathrm{T}}> E_0\cr},\label{wide res}
\end{equation}
the resonance is said to be a wide resonance (this case will be labelled as ``R''). The
$^{12}\mathrm{C}(p,\gamma)^{13}\mathrm{N}$ reaction inside the Sun belongs to this second scenario.

\subsection{Narrow resonances at low energy (MB,r)}
If the conditions in eq.(\ref{narrow res low energy}) hold, the approximate reaction rate reads~\cite{Clayton}
\begin{eqnarray}
r_{ij}^{\mathrm{MB,r}}&=&(2\pi)^{3/2}\frac{N_i N_j}{1+\delta_{ij}}\frac{\omega_{ij}\hbar^2}
{(\mu_{ij}\kb T)^{3/2}}\nonumber\\
&\times&\frac{\Gamma_{\mathrm{in}}(E_{\mathrm{R}})\Gamma_{\mathrm{out}}}{\Gamma_{\mathrm{T}}}
\exp\left(-\frac{E_{\mathrm{R}}}{\kb T}\right), \label{result narrow MB}
\end{eqnarray}
where $\Gamma_{\mathrm{in}}(E_{\mathrm{R}})$ is the entrance channel width, $\Gamma_{\mathrm{out}}$ is the exit channel
width, $\Gamma_{\mathrm{T}}\sim\Gamma_{\mathrm{in}}+\Gamma_{\mathrm{out}}$ is the total resonance width, $\kb T$ is the
plasma thermal energy and $\mu_{ij}$ is the reduced mass of the two-body system $i+j$. In eq.(\ref{result narrow MB}),
the quantum factor $\omega_{ij}$ is defined by~\cite{Blatt}
\begin{displaymath}
\omega_{ij}=\frac{2J_{\mathrm{C}}+1}{(2J_i+1)(2J_j+1)}\; ,
\end{displaymath}
where $J_{\mathrm{C}}$ is the total quantum angular momentum of the compound nucleus and $J_i$, $J_j$ are the spin
numbers of the interacting nuclei $i$ and $j$.

\subsection{Wide resonances (MB,R)}
If the physical conditions in eq.(\ref{wide res}) hold, the approximate reaction rate now reads~\cite{Clayton}
\begin{eqnarray}
r_{ij}^{\mathrm{MB,R}}&=&\frac{2^{3/2}\pi}{3^{1/2}}\frac{N_i N_j}{1+\delta_{ij}}\frac{\omega_{ij}\hbar^2E_0^{1/2}}
{\mu_{ij}^{3/2}\kb T}\nonumber\\
&\times& \frac{\Gamma_{\mathrm{in}}\Gamma_{\mathrm{out}}}{(E_0-E_{\mathrm{R}})^2+(\Gamma_{\mathrm{T}}/2)^2}
\exp\left(-\frac{3E_0}{\kb T}\right).\label{result wide MB}
\end{eqnarray}

Equation (\ref{result wide MB}) has been obtained by writing the astrophysical factor $S_{ij}$ as
\begin{equation}
S_{ij}(E)=\frac{\pi}{2}\frac{\hbar^2}{\mu_{ij}}\frac{\omega_{ij}\Gamma_{\mathrm{in}}\Gamma_{\mathrm{out}}}
{(E-E_{\mathrm{R}})^2+(\Gamma_{\mathrm{T}}/2)^2}\;,\label{astro factor}
\end{equation}
through a well known Breit-Wigner approximation formula, which is supposed to hold if the energy $E$ is close to the
resonance energy $E_{\mathrm{R}}$.

\section{Non-extensive resonant reaction rates}\label{nonext}
In refs.~\cite{LavQua,KLLQ,CKLLMQ} it has been shown that a coherent model describing a given stellar core might deal
with slightly coupled plasmas, non-Markovian random walks, many-body collisions and memory effects. From this point of
view, the classical theory founded on the Maxwell-Boltzmann statistical mechanics should be considered a first order
approximation (a very good one indeed inside the Sun's core). A further approximation, to which we are referring from
now on, lies on the more general picture of non-extensive statistical mechanics~\cite{Tsallis,GellMann}.

Here we limit ourselves to briefly outline, by means of three different approaches, how non-standard statistics can
arise. In Fokker-Planck equation context, we can introduce corrections to the friction and diffusion coefficients,
considering their expressions to the next order in the velocity variable; stationary solution is the Tsallis
non-extensive distribution. In a plasma, each particle is affected by the total electric field distribution of the
other charges; the total microfields have a relatively small random component, generally show long time correlations
and generate anomalous diffusion. All these effects imply a deviation from the Maxwell-Boltzmann distribution whose
entity depends on the plasma parameter and on an ion-ion correlation parameter. Correlations among the collective
modes can lead to a long time asymptotic behavior of the velocity correlations of the ions and to anomalous
diffusion, that are related to generalized entropy and generate non-Maxwellian probability distributions.

In this new picture, the reaction rates are defined in the same way as in eq.(\ref{reaction rate def}), but now the
non-extensive energy distribution function $\phi_{\mathrm{NE}}(E)$ is used, instead of the usual Maxwell-Boltzmann
distribution $\phi_{\mathrm{MB}}(E)$.

It has been shown~\cite{KLLQ,Clayton paper,CKLLMQ} that the analytical relationship linking $\phi_{\mathrm{NE}}$ and
$\phi_{\mathrm{MB}}$ can be cast, to first order of approximation, in the following fashion
\begin{equation}
\phi_{\mathrm{NE}}(E)=\left(1+\frac{15}{4}\delta\right)\phi_{\mathrm{MB}}(E)\exp\left[-\delta\left(\frac{E}{\kb
T}\right)^2\right],\label{nonext distribution}
\end{equation}
provided that $|\delta|\ll\kb T/E$ (or $|\delta|\ll\kb T/E_0$, for computational purposes). The order of magnitude of
the deformation parameter $\delta$ is $10^{-3}\div 10^{-2}$ inside the Sun, but it can be higher in extremely dense
stellar plasmas ({\it e.g.} in white dwarfs). The $\delta$ parameter is linearly related to the $q$ entropic parameter
that appears in the non-extensive formalism~\cite{Tsallis}: the relationship between them two is $\delta=(1-q)/2$.

We can distinguish two physically relevant cases: if $\delta<0$, the high energy tail of the $\phi_{\mathrm{NE}}$
distribution function is increased with respect to $\phi_{\mathrm{MB}}$ (super-extensivity), while if $\delta>0$, the
high energy tail is depleted (sub-extensivity).

Now we are able to develop the entire non-extensive resonant formalism of the thermonuclear reaction rates.

\subsection{Narrow resonances at low energy (NE,r)}\label{narrow NE r}
From eqs.(\ref{reaction rate def}) and (\ref{nonext distribution}), the non-extensive rate of a given reaction proceeding
through a narrow resonance at low energy reads
\begin{eqnarray}
r_{ij}^{\mathrm{NE,r}}&=&\frac{N_i N_j}{1+\delta_{ij}}\left(1+\frac{15}{4}\delta\right)\int_{0}^{+\infty}
\phi_{\mathrm{MB}}(E)\nonumber\\
&\times& v_{ij}(E)\sigma_{ij}(E)\exp\left[-\delta\left(\frac{E}{\kb T}\right)^2\right]\mathrm{d}E\;.\label{rate NE r}
\end{eqnarray}

Starting from the hypotheses in eq.(\ref{narrow res low energy}) we can state that, in the energy interval
$E_{\mathrm{R}}-\Gamma_{\mathrm{T}}<E<E_{\mathrm{R}}+\Gamma_{\mathrm{T}}$, the integral in eq.(\ref{rate NE r}) is
strongly ruled by the reaction cross section function $\sigma_{ij}(E)$ only. Thus we can write, with very good
approximation, the following result
\begin{equation}
r_{ij}^{\mathrm{NE,r}}=r_{ij}^{\mathrm{MB,r}}\left(1+\frac{15}{4}\delta\right)
\exp\left[-\delta\left(\frac{E_{\mathrm{R}}}{\kb T}\right)^2\right]\;.\label{rate NE r final}
\end{equation}

If $|\delta|\ll (\kb T/E_{\mathrm{R}})^2$, we can linearize eq.(\ref{rate NE r final}), and the first order formula is
\begin{equation}
r_{ij}^{\mathrm{NE,r}}=r_{ij}^{\mathrm{MB,r}}[1+C_1(\kb T,E_{\mathrm{R}})\delta]\;,\label{rate NE r linearized}
\end{equation}
where
\begin{displaymath}
C_1(\kb T,E_{\mathrm{R}})=\frac{15}{4}-\left(\frac{E_{\mathrm{R}}}{\kb T}\right)^2.
\end{displaymath}

Therefore, our final result in the case of narrow resonances at low energy is expressed by
\begin{equation}
r_{ij}^{\mathrm{NE,r}}=r_{ij}^{\mathrm{MB,r}}\left[1+\frac{15}{4}\delta-\left(\frac{E_{\mathrm{R}}} {\kb
T}\right)^2\delta\right]\; . \label{narrowres NE}
\end{equation}

If the condition
\begin{equation}
\frac{E_{\mathrm{R}}}{\kb T}\approx\frac{E_0}{\kb T}>\sqrt{\frac{15}{4}}\simeq 1.936\; ,  \label{condition}
\end{equation}
is satisfied, then $C_1(\kb T,E_{\mathrm{R}})<0$. In the Sun interior, the thermal energy is $\kb T\simeq
1.36\,\mathrm{keV}$ and $E_0$ is in the energy interval $24\,\mathrm{keV}<E_0<29\,\mathrm{keV}$ for the CNO cycle, and
$E_0\approx 6\,\mathrm{keV}$ for the $p+p\rightarrow d+e^+ +\nu_e$ reaction. Thus, in the Sun's core, and in many
other stellar plasmas of interest, eq.(\ref{condition}) is always satisfied; then, from eq.(\ref{rate NE r linearized}),
we can actually state that the non-extensive rate is increased or diminished with respect to the classically
calculated rate, whether $\delta<0$ or $\delta>0$ (as already mentioned in sect.~\ref{nonext}).

\subsection{Wide resonances (NE,R)}\label{wide NE R}
In the case of a generic reaction (no matter if resonant or not), the following formula linking $r_{ij}^{\mathrm{NE}}$
and $r_{ij}^{\mathrm{MB}}$ holds,
\begin{equation}
r_{ij}^{\mathrm{NE}}=r_{ij}^{\mathrm{MB}}\frac{S_{ij}(\tilde{E}_0)}{S_{ij}(E_0)}
\left(1+\frac{15}{4}\delta-\frac{7}{3}\delta\frac{E_0}{\kb T}\right)\exp(-\Delta_{ij})\;,\label{first}
\end{equation}
where
\begin{eqnarray}
\Delta_{ij}&\equiv&\Delta_{ij}(\delta,\tilde{E}_0)=-\frac{3E_0}{\kb T}\nonumber\\
&\times&\left[1-\left(1+\frac{5}{3}\delta\frac{\tilde{E}_0}{\kb T}\right) \left(1+2\delta\frac{\tilde{E}_0}{\kb
T}\right)^{-2/3}\right],\label{second}
\end{eqnarray}
and
\begin{equation}
\tilde{E}_0=E_0\left(1+2\delta\frac{\tilde{E}_0}{\kb T}\right)^{-2/3}.\label{ezerop}
\end{equation}

Equations (\ref{first}) and (\ref{second}) express the first order correction to the classical reaction rate
$r_{ij}^{\mathrm{MB}}$, due to the deformed distribution function that has been already defined in eq.(\ref{nonext
distribution}). Besides, eq.(\ref{ezerop}) shows implicitly the relationship between the new (deformed) Gamow energy
$\tilde{E}_0$, and the classical one $E_0$. A complete proof of these equations can be found in ref.~\cite{Clayton
paper}, in which the authors started from an \textit{ad hoc} assumption of the deformed distribution function,
$\phi_{\mathrm{NE}}(E)$, regardless of any statistical basis. In this paper, we adopt the same results, but lying on the
ground of non-extensive statistical mechanics (as was outlined, for example, in ref.~\cite{CKLLMQ}).

In eq.(\ref{first}), the physical properties of the reaction are summarized in the ratio between the astrophysical
factor at the new Gamow energy, $S_{ij}(\tilde{E}_0)$, and the same factor at $E_0$ energy, $S_{ij}(E_0)$. Our aim is
to express that ratio as a function of $E_{\mathrm{R}}$ and $\Gamma_{\mathrm{T}}$ only. The deformation parameter
$\delta$ is assumed to satisfy the following inequality
\begin{equation}
|\delta|\ll\frac{\kb T}{E_0}\; .  \label{hypothesis}
\end{equation}

Then we can write, as a formal expansion in $\delta$,
\begin{equation}
S_{ij}(\tilde{E}_0)\approx S_0+S_1\delta\; \label{first order astro factor}
\end{equation}
where the two coefficients $S_0$ and $S_1$ can be calculated as follows. From eq.(\ref{astro factor}), we immediately
obtain that
\begin{equation}
S_0=S_{ij}(E_0)=\frac{\pi}{2}\frac{\hbar^2}{\mu_{ij}}
\frac{\omega_{ij}\Gamma_{\mathrm{in}}\Gamma_{\mathrm{out}}}{(E_0-E_{\mathrm{R}})^2+(\Gamma_{\mathrm{T}}/2)^2} \; .
\label{s0}
\end{equation}

On the contrary, the first order coefficient in eq.(\ref{first order astro factor}) reads
\begin{equation}
S_1=-\pi\omega_{ij}\frac{\hbar^2}{\mu_{ij}}\left.\frac{(E_0-E_{\mathrm{R}})\Gamma_{\mathrm{in}}\Gamma_{\mathrm{out}}}
{[(E_0-E_{\mathrm{R}})^2+\Gamma_{\mathrm{T}}^2/4]^2}\frac{\mathrm{d}\tilde{E}_0}{\mathrm{d}\delta}\right|_{\delta=0}\;
. \label{s1}
\end{equation}

In order to calculate the analytical expression for the $\mathrm{d}\tilde{E}_0/\mathrm{d}\delta|_{\delta=0}$
derivative, we differentiate both members of eq.(\ref{ezerop}) with respect to $\delta$, obtaining
\begin{displaymath}
\frac{\mathrm{d}\tilde{E}_0}{\mathrm{d}\delta}=-\frac{4}{3}\frac{E_0}{\kb T}
\left(\tilde{E}_0+\frac{\mathrm{d}\tilde{E}_0}{\mathrm{d}\delta}\delta\right)\left(1+2\frac{\tilde{E}_0}
{\mathrm{k_B}T}\delta\right)^{-5/3},
\end{displaymath}
from which, without any approximation, it follows that
\begin{equation}
\left.\frac{\mathrm{d}\tilde{E}_0}{\mathrm{d}\delta}\right|_{\delta=0}=-\frac{4}{3}\frac{E_0^2}{\kb T}\; .
\label{derivative}
\end{equation}
Now we can rewrite eq.(\ref{first order astro factor}), using the results of eqs.(\ref{s0}) and (\ref{derivative}), as
\begin{displaymath}
S_{ij}(\tilde{E}_0)\approx S_{ij}(E_0)\left(1+\frac{S_1}{S_0}\delta\right)
\end{displaymath}
and then, the $S_{ij}(\tilde{E}_0)/S_{ij}(E_0)$ ratio eventually becomes
\begin{equation}
\frac{S_{ij}(\tilde{E}_0)}{S_{ij}(E_0)}=1+\frac{8}{3}\delta\frac{E_0-E_{\mathrm{R}}}{(E_0-E_{\mathrm{R}})^2+
\Gamma_{\mathrm{T}}^2/4}\frac{E_0^2}{\kb T}\;.\label{ratio}
\end{equation}

Placing the previous result of eq.(\ref{ratio}) into eq.(\ref{first}), it is clear that the non-extensive reaction rate,
to first order, reads
\begin{eqnarray}
r_{ij}^{\mathrm{NE,R}}&=&r_{ij}^{\mathrm{MB,R}}\exp(-\Delta_{ij})\left[1+\frac{15}{4}\delta-\frac{7}{3}\delta
\frac{E_0}{\kb T}+\right.\nonumber\\
&+&\left.\frac{8}{3}\delta\frac{(E_0-E_{\mathrm{R}})E_0}{(E_0-E_{\mathrm{R}})^2+\Gamma_{\mathrm{T}}^2/4}\frac{E_0}{\kb
T}\right]\; . \label{result}
\end{eqnarray}

A further approximation, under the hypothesis stated in eq.(\ref{hypothesis}), is
\begin{displaymath}
\exp(-\Delta_{ij})\approx 1-\left(\frac{E_0}{\kb T}\right)^2\delta\; ,
\end{displaymath}
and therefore, from eq.(\ref{result}), the final result immediately follows
\begin{eqnarray}
r_{ij}^{\mathrm{NE,R}}&=& r_{ij}^{\mathrm{MB,R}}\left[1+\frac{15}{4}\delta-\frac{7}{3}\delta \frac{E_0}{\kb T}-\delta
\left(\frac{E_0}{\kb T}\right)^2+\right.\nonumber\\
& &+\left.\frac{8}{3}\delta\frac{(E_0-E_{\mathrm{R}})E_0}{(E_0-E_{\mathrm{R}})^2+\Gamma_{\mathrm{T}}^2/4}\frac{E_0}{\kb
T}\right]\;.\label{rate NE wideres}
\end{eqnarray}

\section{Conclusions and discussion}
In this work we have analytically derived two first order formulae that can be used to express the non-extensive
reaction rate as a product of the classical reaction rate times a suitable corrective factor for both narrow and wide
resonances, as shown in eqs.(\ref{narrowres NE}) and (\ref{rate NE wideres}). It should be stressed that the previous
results are correct only if $|\delta|$ is very small (in the sense of eq.(\ref{hypothesis})), namely if we are dealing
with slight deformations of the energy distribution function. This is really the most common situation, as far as
astrophysical plasmas are concerned (in fact $|\delta|\sim 10^{-3}\div 10^{-2}$).

Concerning the fusion reactions between two medium-weighted nuclei, for example the $^{12}\mathrm{C}+{^{12}\mathrm{C}}$
reaction, our non-extensive factor, which now can be formally defined as follows,
\begin{displaymath}
f_{\mathrm{NE}}=1+\frac{15}{4}\delta-\left(\frac{E_{\mathrm{R}}}{\mathrm{k_B} T}\right)^2\delta\; ,
\end{displaymath}
gives rise to further correction beside the screening and the potential resonant screening, already investigated
in~\cite{Langanke} and~\cite{Japan}.

It is important to point out that the non-extensivity does not affect the other plasma corrections: therefore, we can
define an effective factor $F$ as
\begin{equation}
F=f_{\mathrm{NE}}\cdot f_{\mathrm{S}}\cdot f_{\mathrm{RS}}\;,\label{final corrective factor}
\end{equation}
where $f_{\mathrm{S}}$ and $f_{\mathrm{RS}}$ account for the \debhu\/ screening and the resonant screening effect
respectively.

We have applied our results to a physical model describing a carbon white dwarf's plasma, with a temperature of
$T=8\cdot 10^8\,\mathrm{K}$ and a mass density of $\rho=2\cdot 10^9\,\mathrm{g\cdot cm^{-3}}$ (the plasma parameter is,
correspondingly, $\Gamma\simeq 5.6$). Furthermore, we have set a deformation parameter $|\delta|=10^{-3}$, regardless
of its sign, and we have kept the energy of the possible resonance, $E_{\mathrm{R}}$, as a free parameter.

\begin{figure}[t]

\includegraphics[width=.45\textwidth,height=.4\textwidth]{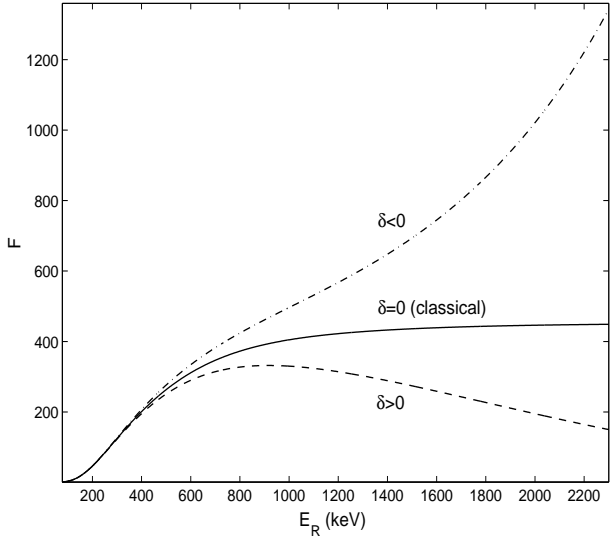}

\caption{Linear plot of the effective factor $F$, defined in eq.(\ref{final corrective factor}), against the resonance
energy $E_{\mathrm{R}}$. The dash-dotted (upper) line refers to super-extensivity, the dashed (lower) line to
sub-extensivity, while the solid (middle) line describes the classical (MB) result.} \label{F-factor figure}

\end{figure}

In fig.~\ref{F-factor figure}, we plot the total corrective factor defined in eq.(\ref{final corrective factor}) 
against the energy at which the narrow resonance is supposed to be located, in the range between $0\,\mathrm{keV}$ and
$2400\,\mathrm{keV}$, considering three very different conditions (depending on the sign and value of the deformation
parameter $\delta$). In our calculations, in order to estimate the functional dependence of $f_{\mathrm{S}}\cdot
f_{\mathrm{RS}}$ with respect to $E_{\mathrm{R}}$, we have adopted the fitting formulae derived by Itoh and
collaborators~\cite{Japan}: their result, that has been derived through a classical treatment, is also shown in
fig.~\ref{F-factor figure} by the line labelled with $\delta=0$. From the same figure, it is clear that a slight
non-extensivity does introduce non-trivial corrections that become more and more important as the resonance energy
rises: if $\delta>0$, our total effective factor is $F\simeq 136$ at $E_{\mathrm{R}}=2.4\,\mathrm{MeV}$ while, at the
same resonance energy, the factor is $F\simeq 1484$ if $\delta<0$. Anyway, in both cases, $\lim F=1$ when
$E_{\mathrm{R}}\rightarrow 0\,\mathrm{keV}$. It is also noteworthy that the effective factor always acts to enhance the
resonant reaction rate, no matter if the system is super or sub-extensive. In conclusion, all the plasma enhancements
due to the presence of long range many-body nuclear correlations and memory effects, that can be described within the
non-extensive statistics by means of the entropic parameter $q>1$ ($\delta<0$), are in the direction of still more
increasing the effective factor $F$ of nuclear rates of hydrostatic burning and white dwarfs environment.


\end{document}